\documentclass[aps,pra,showpacs,groupedaddress,twocolumn,amsmath,amssymb,nofootinbib]{revtex4-1}

\usepackage[utf8]{inputenc} \usepackage[american]{babel}
\usepackage{csquotes} 
\usepackage{graphicx}
\usepackage{color} \usepackage{grffile} \usepackage{units}
\usepackage{amsfonts}

\newcommand{\ket}[1]{\left|#1\right\rangle}
\newcommand{\bra}[1]{\left\langle #1\right|}

\newcommand{\var}{\mathrm{var}}
\newcommand{\ii}{\mathrm{i}}

\begin{document}

\title{Atomic nonclassicality quasiprobabilities}
	
\author{T. Kiesel and W. Vogel}

\affiliation{Arbeitsgruppe Quantenoptik, Institut für Physik,
  Universität Rostock, D-18051 Rostock, Germany}

\author{S. L. Christensen, J.-B. Béguin, J. Appel, E. S.  Polzik}
\affiliation{QUANTOP, Niels Bohr Institute, University of Copenhagen,
  Blegdamsvej 17, 2100 K\o{}benhavn \O, Denmark}

\begin{abstract}
  Although nonclassical quantum states are important both conceptually
  and as a resource for quantum technology, it is often difficult to
  test whether a given quantum system displays nonclassicality. A
  simple method to certify nonclassicality is introduced, based on
  easily accessible collective atomic quadrature measurements, without
  the need of full state tomography. The statistics is analyzed beyond
  the ground-state noise level, by direct sampling of a regularized
  atomic quadrature quasiprobability. Nonclassicality of a squeezed
  ensemble of $2\cdot 10^5$~Cesium atoms is demonstrated, with a
  significance of up to $23$ standard deviations.
\end{abstract}


%03.65.Ta 	Foundations of quantum mechanics; measurement theory (for optical tests of quantum theory, see 42.50.Xa)
%42.50.Ct 	Quantum description of interaction of light and matter; related experiments 
%42.50.Dv 	Quantum state engineering and measurements (see also 03.65.Ud Entanglement and quantum nonlocality, e.g., EPR paradox, Bells inequalities, GHZ states, etc.)
%42.50.Xa 	Optical tests of quantum theory 
%03.70.+k, Theory of quantized fields (see also 11.10.-z Field theory)
%42.50.Lc 	Quantum fluctuations, quantum noise, and quantum jumps
%75.10.Jm 	Quantized spin models, including quantum spin frustration 
\maketitle

\section{Introduction}
The difference between the classical and quantum mechanical
description of physical systems is often discussed in terms of
nonclassicality. This topic is of general interest, since it gives
insight into the reasons why a classical picture of nature has to
fail, and provides methods to explore what a quantum mechanical system
can do which a classical system cannot.  In particular, such issues
are of great importance for numerous fields in quantum optics, 
quantum computation, quantum information, and related research areas.

Nonclassicality of the harmonic oscillator has been well understood
during the last years. Its definition is based on the coherent states
$\ket\alpha$, parameterized by a complex amplitude $\alpha$.  The
evolution of these pure quantum states most closely corresponds to the
classical evolution of position and momentum.  They are minimum
uncertainty states with a stable Gaussian wave packet. In this sense
they represent the closest analogue to the classical behavior.
Furthermore, they allow to decompose any quantum state in the form
\begin{equation}
  \hat{\rho} = \int d^2\alpha\, P(\alpha) \ket{\alpha}\bra{\alpha},
\label{eq:GS}
\end{equation}
where $P(\alpha)$ is the Glauber-Sudarshan $P$
function~\cite{PhysRevLett.10.277, PhysRev.131.2766}. In this
representation, an arbitrary quantum state formally resembles a
statistical mixture of (classical) coherent states. However,
for many states $P(\alpha)$ may show negativities, which cannot be
interpreted in terms of statistical probability densities. Such states
are consequently referred to as nonclassical~\cite{TitulaerGlauber}.

Atomic coherent states (ACS) have already been developed in
\cite{arecchi1972atomic}. It has been shown that their interaction
with a classical electric and magnetic field can be understood
classically, cf.~\cite{perelomov72}, and any density operator can be
formally written as a statistical average over ACS. This remarkable
property allows one to define nonclassicality in a way closely
corresponding to that of the harmonic
oscillator~\cite{luis2006nonclassical,Giraud2008}. Specific
nonclassical properties of atomic systems have already been discussed in previous publications, such as spin squeezing~\cite{kitagawa1993squeezed,Kuzmich1997,Kuzmich1998} and squeezing in Ramsey spectroscopy~\cite{wineland1992spin}. The creation of entanglement in spin systems~\cite{Sorensen2001,RMP2010} and the relation between entanglement and squeezing~\cite{luis2006polarization} have also been studied. Recently, nonclassicality has been examined in terms of measurable probability distributions~\cite{luis2011angular}.

The present manuscript addresses a method to visualize
nonclassicality of atomic systems, characterized by angular momentum
states with a large orientation in the $z$-direction and small
fluctuations along the two orthogonal directions. 
To provide a
simple and experimentally accessible condition for the detection of
the nonclassical features of the system, we introduce the concept of a
regularized atomic quadrature quasiprobability (AQQP). 
Our method is applied to
ex\-peri\-men\-tal data from approximately $10^5$ two-level atoms,
which were collectively prepared in an atomic spin-squeezed state. It
uncovers the nonclassical properties, even for weak
squeezing.

The paper is organized as follows: In Sec.~II, we give a short introduction into the quantum mechanical description of the angular momentum, and review the Holstein-Primakoff approximation. Then, we explain the verification of nonclassical effects by nonclassicality quasiprobabilities and define nonclassicality quadrature distributions. The experimental results, including the setup and data analysis, are given in Sec.~III. Finally, we summarize our conclusions in Sec.~IV.

\section{Atomic coherent states}
\subsection{Basic relations}

Formally, the atomic ensemble under study can be characterized as an
angular momentum system.  In quantum mechanics, the angular momentum
is described by a vector valued operator $\hat {\vec J} = (\hat
J_x,\hat J_y,\hat J_z)$, satisfying the commutation relations (setting
$\hbar=1$):
\begin{equation}
  [\hat J_x,\hat J_y] = \ii \hat J_z,\quad [\hat J_y,\hat J_z] = \ii \hat J_x,\quad [\hat J_z,\hat J_x] = \ii\hat J_y.
\end{equation}
Furthermore, we introduce the angular momentum ladder operators $\hat
J_\pm = \hat J_x \pm \ii \hat J_y$, whose commutator is given by $ [\hat
J_+,\hat J_-] = 2 \hat J_z$.  Throughout this work, we keep the
quantum number $j$ fixed, while the quantum number $m$ takes values in
$\{-j,-j+1,\ldots,j\}$.

The foundation for the discussion of nonclassicality is typically
given by the coherent states, being defined as~\cite{Perelomov1986}
\begin{equation}
  \ket{\theta,\varphi} = \exp\{(\theta e^{\ii\varphi}\hat J_- - \theta e^{-\ii\varphi}\hat J_+)/2\} \ket{j;j},
\end{equation}
where the fiducial state $\ket{j;j}$ is the common eigenstate of the
operators $\hat J^2$ and $\hat J_z$ with maximum eigenvalue for the
latter operator. These coherent states can be seen as the analog to a
classical angular momentum state. They have minimum uncertainty in
$\hat J^2$, and their expectation value of the projection of
$\hat{\vec J}$,
\begin{equation}
  \vec n(\theta,\varphi) \cdot \hat{\vec J} =
  \sin(\theta)\cos(\varphi) \hat J_x + \sin(\theta)\sin(\varphi) \hat
  J_y + \cos(\theta) \hat J_z\label{eq:J:theta:phi}
\end{equation}
is exactly equal to $j$, without uncertainty, such that the state
$\ket{\theta,\varphi}$ can be understood as pointing exactly in the
direction of $\vec n(\theta,\varphi)$. Moreover, their time evolution
in a classical magnetic field exactly matches the classical
evolution~\cite{perelomov72}. Finally, there exists a representation
of the density operator of an arbitrary quantum state in the form
\begin{equation}
  \hat\rho = \frac{2 j + 1}{4\pi} \int_0^\pi \sin \theta d\theta\int_0^{2\pi} d\varphi \, P(\theta,\varphi) \ket{\theta,\varphi}\bra{\theta,\varphi}.
\end{equation}
Similar to Eq.~(\ref{eq:GS}), this is formally equivalent to a
decomposition of a quantum state into a statistical mixture of
(classical) coherent states. Therefore, all quantum states, for which
the weight function $P(\theta,\varphi)$ is non-negative, are classical
angular momentum states. Conversely, any quantum state for which such
a non-negative function $P(\theta,\varphi)$ does not exist, will be
referred as being nonclassical~\cite{luis2006nonclassical,Giraud2008}.

\subsection{The Holstein-Primakoff-approximation}

Let us consider a system with a large angular momentum, i.e.~$j \gg
1$, which is mainly aligned in the $z$-direction. In this case, the
operator $\hat J_z$ can be approximated by its expectation value
$\langle\hat J_z\rangle\approx j$, being a real number as in classical
physics.  Only the remaining angular momentum components $\hat J_x,
\hat J_y$ are treated fully quantum-mechanically~\cite{holstein1940field}. Then, the ladder
operators satisfy the approximated commutator relation
\begin{equation}
  [\hat J_+,\hat J_-] \approx 2 j.
\end{equation}

The re-normalized operators $\hat j_\pm = \frac{1}{\sqrt{2j}}\hat
J_\pm$ obey the same algebra as the bosonic operators.  We may
introduce rescaled angular momentum components via $\hat j_\pm =
\frac{1}{2}(\hat j_x \pm \ii \hat j_y)$, which are defined such that the
uncertainty of the ground state, $\bra{j;j}(\Delta \hat
j_x)^2)\ket{j;j}= \bra{j;j}(\Delta \hat j_y)^2)\ket{j;j}=1$. The
weight function $P(\theta,\varphi)$ corresponds to the
Glauber-Sudarshan $P$ function of the quantum state, with the complex
argument $\alpha = \sqrt{j/2} \theta e^{\ii\varphi}$.  The properties of
$P(\alpha)$ form the basis for defining nonclassicality of the
harmonic oscillator. Consequently, we can reformulate established
  methods for nonclassicality for angular momentum
  systems.

\subsection{Atomic nonclassicality quasiprobabilities}

Although the $P$ function uniquely defines the nonclassicality of a
quantum system, its reconstruction is in general impossible for the harmonic
oscillator, since it can be highly singular. For
getting definite answers about nonclassical effects, one may introduce a
filtered atomic $P$ function, in close analogy to that of the harmonic oscillator~\cite{Kiesel2010}.
For atomic systems, one multiplies the characteristic function $\Phi(\xi)$
of $P(\theta,\varphi)$, 
\begin{equation}
  \Phi(\xi) \equiv  \langle e^{\xi\hat j_- - \xi^*\hat j_+} \rangle e^{|\xi|^2/2},\label{eq:char:func}
\end{equation}
with a suitable filter function $\Omega_w(\xi)$ (parameterized by $w$),
\begin{equation}
  \Phi_\Omega(\xi) \equiv \Phi(\xi) \Omega_w(\xi).
\end{equation}
The atomic nonclassicality quasiprobability is then given by
\begin{equation}
  P_\Omega(j_x,j_y) \equiv \frac{1}{\pi^2}\int_\mathbb{C} d^2\xi \, e^{\ii (j_y \xi_r - j_x \xi_i)} \Phi_\Omega(\xi),
\end{equation}
with $\xi=\xi_r + \ii \xi_i$. The filter function $\Omega_w(\xi)$ has to
satisfy certain properties:
First, it has to guarantee that the nonclassicality quasiprobability
of any state is always regular. Second, it only should show
negativities if the quantum state is nonclassical. For the harmonic
oscillator we have shown~\cite{Kiesel2010} that for any nonclassical
state one finds a nonclassicality quasiprobability demonstrating
negativities by simply varying a real parameter $w$ that scales the
filter width.

\subsection{Atomic quadrature quasiprobability}
In the following we will introduce the AQQP to efficiently characterize 
nonclassicality of atomic spin systems with a minimum of measured data. Quadrature distributions are just probability densities, so that a regularization procedure is superfluous. The situation changes if 
one analyzes the quadrature statistics beyond the ground-state noise level~\cite{Vogel2000}, to visualize usually hidden  nonclassical effects, also beyond the negativities of the Wigner function~\cite{Lv-Sh}. Subsequent filtering yields the marginals of $P_\Omega(j_x,j_y)$.

Introducing atomic quadrature operators $\hat j_\varphi$,
\begin{equation}
\hat{j}_\varphi \equiv \hat{j}_+ e^{\ii \varphi} + \hat{j}_- e^{-\ii \varphi},
\end{equation}
we define the AQQP via
\begin{equation}
  p_\Omega(j_\varphi) \equiv \frac{1}{2\pi}\int_{-\infty}^{\infty} d\lambda \, \Phi(\lambda e^{\ii\varphi}) \, \Omega_w(\lambda) \, e^{\ii \lambda j_\varphi},
\end{equation}
which is just a marginal of $P_\Omega(j_x,j_y)$ for a fixed angle
$\varphi$.  Clearly, if such a marginal distribution shows
negativities, then $P_\Omega(j_x,j_y)$ has negativities as well, and
the state is nonclassical. In general, the inverse conclusion is not
true.

The AQQP distribution can be sampled from a set of $N$ measured
angular momentum values $\{{\tilde j}_k\}_{k=1}^N$ directly by using a
similar approach to~\cite{Kiesel2011}. Hence an estimator for the AQQP
is expressed as an empirical mean of a suitable pattern function
$f_\Omega(\tilde j; j_\varphi, w)$,
\begin{equation}
  p_\Omega(j_\varphi) = \frac{1}{N}\sum_{k=1}^N f_\Omega({\tilde j}_k; j_\varphi, w).\label{eq:estimate:p:Omega}
\end{equation}
The required pattern function is given by
\begin{equation}
  f_\Omega(\tilde j; j_\varphi,w) = \frac{1}{2\pi} \int_{-\infty}^\infty e^{k^2/2} e^{\ii k (\tilde j - j_\varphi)} \Omega_w(k) dk.
\end{equation}
Furthermore, the uncertainty of $p_\Omega(j_\varphi)$ can be estimated
as $1/\sqrt{N}$ times the empirical standard
deviation of the sampled numbers $\{f_\Omega(\tilde
j_k;j_\varphi,w)\}_{k=1}^N$. Therefore, the sampling method is a very
simple way to estimate the quantity of interest together with its
statistical error.

\section{Experimental procedure}
\subsection{Experimental Setup}

The setup is described in detail in
\cite{Appel2009,Louchet-Chauvet2010}: A cloud of $N_a \approx 2\cdot
10^5$ cold Cesium atoms is prepared by laser cooling in a magneto-optical
trap (MOT) and subsequently transferred into a far off-resonant dipole
trap formed in the $\unit[40]{\mu m}$ waist of a single $\unit[5]{W}$,
$\unit[1064]{nm}$ laser beam.  Employing a sequence of optical pumping
pulses, microwave pulses, (optical) purification pulses, a magnetic
bias field of approximately $\unit[1.5]{Gauss}$, and a resonant
microwave $\pi/2$ pulse, each atom is prepared in the state
$\ket{\psi_0}=(\ket{F\!=\!3,m_F\!=\!0}+\ket{F=\!4\!,m_F=\!0})/\sqrt{2}$.

In a simplified picture we describe each atom as a two-level system
with pseudo-spin $1/2$ and with basis states $\ket{\frac{1}{2},\pm
  \frac{1}{2}} := \frac{1}{\sqrt{2}} \left( \ket{F=3} \pm e^{i
    \omega_{34} t} \ket{F=4} \right)$, with $\hbar \omega_{34}=\Delta
E_\text{hyperfine}$.  The atomic population difference in the two
hyper-fine manifolds $F=3$ and $F=4$ now can be regarded as the
$x-$component of a collective $j=N_a/2$-pseudospin vector. This
corresponds to the sum of the individual pseudo-spins, 
$\ket{j;j} = \bigotimes_{i=1}^{N_a}\ket{\psi_0}_i$, an ACS.

By detecting the differential phase shifts $\phi$ that the atomic
cloud imprints on a bichromatic, off-resonant optical probe
pulse~\cite{Louchet-Chauvet2010} we measure the atomic population
difference $\Delta N$ between the two hyper-fine levels. In this way
we perform a non-destructive quantum non-demolition (QND) measurement
of $\hat j_x$, which projects the atomic state into a squeezed state
around $\zeta \phi_1$. Here $\phi_1$ denotes the outcome of the QND
measurement and the parameter $\zeta$ is a function of the QND
measurement strength, which is directly proportional to the optical
density of the ensemble and the probability of spontaneous photon
emission~\cite{Appel2009}.  A~subsequent verification measurement
$\phi_2$, performed on the same ensemble, shows sub-projection noise
quantum fluctuations of the differential phase shift with respect to
the prediction $\zeta \phi_1$.

\begin{figure}
  \centering
  \includegraphics[keepaspectratio,width=\columnwidth]{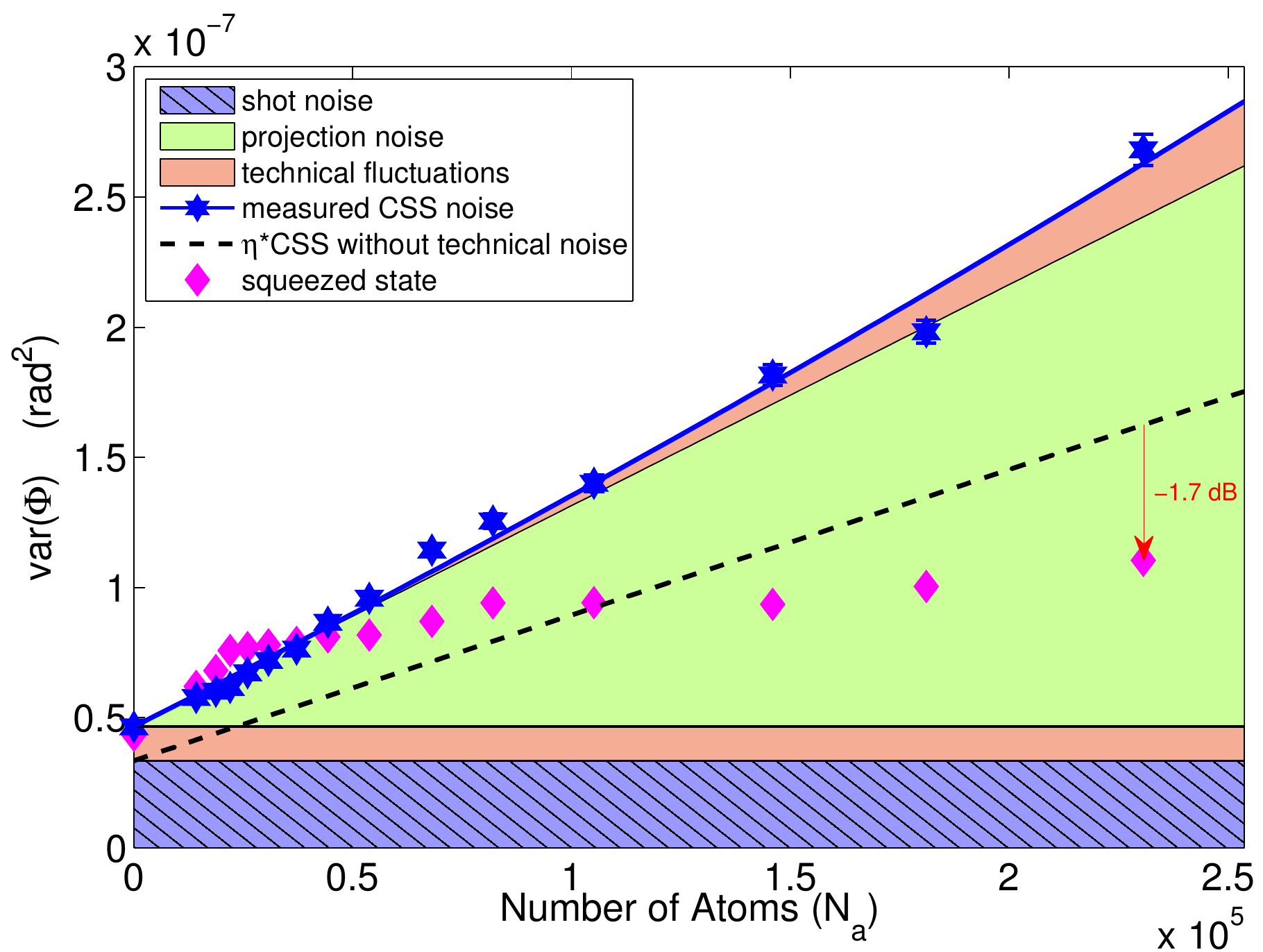}
  \caption{(Color online) Calibration procedure: Blue points: $\var(\bar
    \phi^\text{ACS})$ for different atom numbers; each data point is
    formed by 4841 individual measurements. The scaling of the noise
    as a function of the number of atoms $N_a$ allows us to identify
    noise contributions. Blue (dark shaded) area: light shot noise. Green (bright shaded) area:
    atomic projection noise. Red (grey shaded) areas: technical fluctuations. Black
    dashed line: predicted noise for a ACS with the same collective
    $j_z$ as our squeezed state.  Magenta diamonds:
    $\var({\phi_{12}})$ for experiments with different $N_a$.}
  \label{fig:noise}
\end{figure}

As it is characteristic for QND measurements the signal-to-noise ratio
(and therefore the measurement strength) is limited by the quantum
noise of the meter system -- in our case by the shot noise $\delta n$
of the probe light:
\begin{equation}\phi_1 = \frac{\delta n_1}{n_1} + \kappa \Delta
  N, \qquad \phi_2 = \frac{\delta n_2}{n_2} + \kappa \Delta N,
\end{equation}
where $\kappa$ is the light-atom coupling strength.  Although the
destructive second measurement uses more photons ($n_2 = 1.5\, n_1$)
than the entangling QND measurement, it still contains a fair amount
of light shot noise (see Fig.\ref{fig:noise}).

\subsection{Data analysis}

To be able to convert the $N$ measurement outcomes ${\phi_{12}}_k :=
{\phi_2}_k -\zeta\,{\phi_1}_k$ $(k=1,\dots, N)$ into quadrature
samples, we calibrate our experiment as follows:
\begin{itemize}
\item We prepare a ACS using a varying number of trapped atoms $N_a$
  and, for calibration purposes only, subtract measurement outcomes
  $\phi^\text{ACS}_k$ from successive independent MOT loading cycles
  ($\approx \unit[5]{s}$ apart) to compensate for slow drifts of
  experimental parameters.  By performing a scaling analysis
  \cite{Appel2009} on the differential values $\bar \phi_k^\text{ACS}
  =(\phi^\text{ACS}_k-\phi^\text{ACS}_{k+1})/\sqrt{2}$ we can identify
  the amount of light shot noise $\var(\delta n/n)$ and atomic quantum
  projection noise $\var(\kappa \Delta N)$ and discriminate those
  against technical noise sources.
\item Additionally, due to spontaneous emission processes caused by
  the $\phi_1$-measurement, performed with $n_1=4.1 \cdot 10^7$
  photons, the atomic state is not pure anymore and compared to the
  ACS the Ramsey-fringe contrast (and thus the macroscopic spin $j$ of
  the ensemble) is reduced by a factor $\eta=\exp(-n_1 \epsilon)=65.8\%$,
  where $\epsilon=1.02 \cdot 10^{-8}$ is the decoherence per photon,
  determined experimentally as described in
  \cite{Louchet-Chauvet2010}.
\end{itemize}
With these calibrations we can calculate the phase fluctuations that
we would measure in the absence of any technical noise for a ACS
comprised of $\eta N_a$ atoms, which yields the same Ramsey fringe
amplitude as our squeezed state: \begin{equation}\var_{\text{ACS}(\eta
    N_a)} \equiv \var(\delta n_2/n_2) + \kappa^2 \, \eta \,N_a.
\end{equation}

For a varying number of atoms $0 \le N_a \le 2.9\cdot 10^5$ we perform
in total $N=84730$ experiments and obtain phase shift measurements
${\phi_{12}}_k$, now \emph{without} taking the difference between
subsequent MOT loading cycles.

We convert the ${\phi_{12}}_k$ into quadrature values by normalizing
to $\var_{\text{ACS}(\eta N_a)}$:
\begin{equation}\bar j_k = {\phi_{12}}_k / \sqrt{
    \var_{\text{ACS}(\eta N_a)}}.\end{equation}

The $\bar j_k$ do not only contain the atomic signal $\tilde j_k$ but
also added photon shot noise, which is Gaussian noise, uncorrelated
with the atomic population difference $\Delta N$.  Due to the
normalization procedure the effect of such added Gaussian noise is
equivalent to a detection with a reduced quantum efficiency
\cite{appel07:_electronic_noise_homodyne} (mixing with an ACS). The effective quantum efficiency of our quadrature measurement
depends on the 'atomic signal to shot noise'-ratio and is therefore a
function of $N_a$:
\begin{equation} \text{Efficiency} = \frac{ \var_{\text{ACS}{(\eta
        N_a)}} - \var_{\text{ACS}{(0)}}} {\var_{\text{ACS}{(\eta
        N_a)}} }. \end{equation} In our experiment, it reaches values
up to $83 \%$ for measurements taken with the highest number of atoms
$N_a=2.9 \cdot 10^5$.  This effect only attenuates negativities of the
nonclassicality quasiprobability $P_\Omega$ and any nonclassicality
found in $\bar j_k$ is therefore sufficient to prove a nonclassical
atomic state in the pseudospin operators $ \hat{j}_x, \hat{j}_y$.

\subsection{Verification of nonclassicality}
\begin{figure}
  \includegraphics[width=\columnwidth]{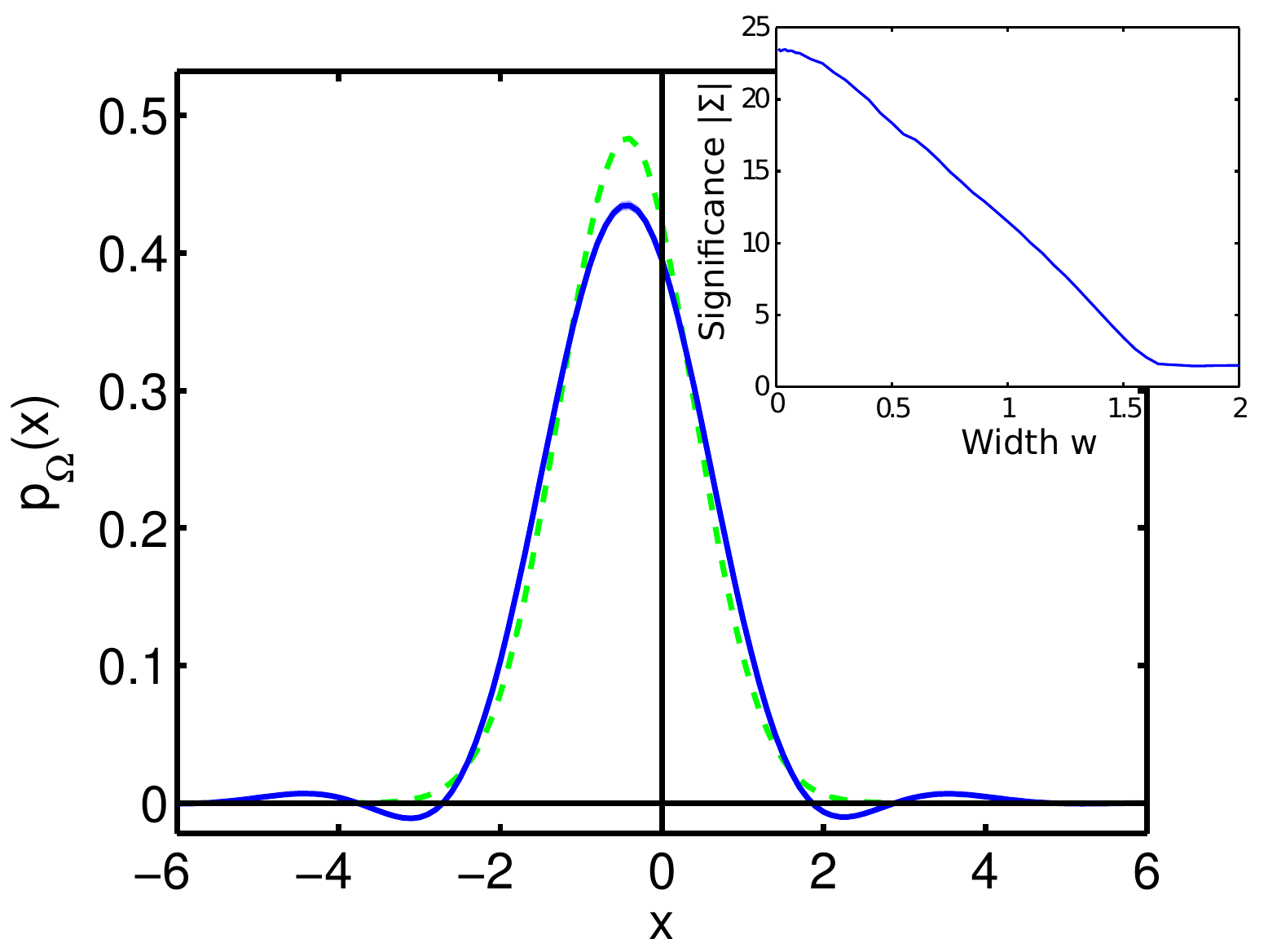}
  \caption{(Color online) AQQP distribution 
    of a squeezed angular momentum state for a filter width $w=1.1$. The error bars are smaller than the linewidths.
    The dashed green curve shows the standard quadrature distribution (i.e.~the
    marginal of the Wigner function), which is non-negative. Inset: Significance $\Sigma$ of the negativity of the AQQP versus the width
    parameter.}
  \label{fig:p:Omega}  \label{fig:sig:w}
\end{figure}

In the following, we only consider experiments performed with an
effective detection efficiency above $77\%$, which show a quadrature
variance $\unit[1.67]{dB}$ below the projection noise.  From the
$\left\{\bar j_k\right\}_{k=1}^{N=4841}$ data points, we can estimate
the AQQP according to
Eq.~(\ref{eq:estimate:p:Omega}), by using a one-dimensional
autocorrelation filter
\begin{equation}
  \Omega_w(k) = \mathcal N^{-1}\int \omega(k')\omega(k'+k/w) dk,
  \label{eq:autocorrelation}
\end{equation}
with $\omega(k) = e^{-k^4}$ and $\mathcal N = \int [\omega(k)]^2 dk$.
This filter preserves all information about the distribution, but only
displays negativities for nonclassical states.  First, we look at
the significance of the negativity as a function of the filter width,
\begin{equation}
  \Sigma(w) = \min\left(\frac{p_\Omega(j_\varphi)}{\sigma(p_\Omega(j_\varphi))}\right),
\end{equation}
where $\sigma(p_\Omega(j_\varphi))$ is the standard deviation of the
estimate $p_\Omega(j_\varphi)$. This quantity is nothing but the value
of the negativity in terms of standard deviations of the measurement.
For observing nonclassical effects, this quantity has to be negative.
From our experimental data, we always observe such negative
significance.  Moreover, the inset in Fig.~\ref{fig:sig:w} shows the
dependence of its absolute value on the filter width $w$.  It can be
seen that there is a filter width, for which the significance is
optimized. In case of squeezing, this may even be arbitrarily close to
$w=0$. Here, we can obtain a significance of up to $23$ standard
deviations.

The main part of Fig.~\ref{fig:p:Omega} shows the AQQP of the
corresponding state. Here, the filter width is chosen to be $w = 1.1$,
which gives a smaller significance of $10.0$ standard deviations, but
also leads to more pronounced effects. We observe oscillations with
significant negativities. Since the autocorrelation function
(Eq.~(\ref{eq:autocorrelation})) fulfills the requirements of a filter
function as listed in the theory part by construction, the
negativities of the AQQP clearly reflect the nonclassicality of the
quantum state.

\section{Conclusions}

We introduced the atomic quadrature quasiprobability,
$p_\Omega(j_\varphi)$, for atomic systems described by angular
momentum states in the Holstein-Primakoff approximation. Its
negativities certify negativities of the Glauber-Sudarshan 
$P$-function of the atomic system, and hence the nonclassicality of
the state beyond both the ground-state noise level and negativities of the Wigner function. We apply our method to experimental data observed in a
squeezed atomic ensemble and demonstrate nonclassicality with a
  statistical significance of up to $23$ standard deviations. We
consider our method a valuable tool for certifying nonclassicality
especially for states prepared in composed many-identical-particle
systems such as cold atomic clouds, vapor cells or rare-earth doped
crystals for quantum technology and -communication applications.

\section*{Acknowledgments}
This work was supported by the Deutsche Forschungsgemeinschaft through
SFB 652, by Danmarks Grundforskningsfond, and by Q-ESSENCE.

\end{document}